\documentclass{PoS}

\def \<{\langle}
\def \>{\rangle}
\def \+{\dagger}
\def \({\left(}
\def \){\right)}

\def \[{\left[}
\def \]{\right]}

\def \min {\rm min}

\def \pinit {p_{\rm in}}
\def \CT {\Delta t}

\def \hard {{\rm hard}}

\title{Finding the Scatterers in Hot Quark Soup}

\ShortTitle{Finding the Scatterers in Hot Quark Soup}

\author{Francesco  D'Eramo\\
        Dipartimento di Fisica e Astronomia, Universit\`a di Padova, Via Marzolo 8, 35131 Padova, Italy\\
        INFN, Sezione di Padova, Via Marzolo 8, 35131 Padova, Italy
        E-mail: \email{francesco.deramo@pd.infn.it}}

\author{Krishna Rajagopal\\
          Center for Theoretical Physics, Massachusetts Institute of Technology, Cambridge, MA 02139, USA\\
        E-mail: \email{krishna@mit.edu}}

\author{\speaker{Yi Yin}
            \\
        Center for Theoretical Physics, Massachusetts Institute of Technology, Cambridge, MA 02139, USA\\
        E-mail: \email{yiyin3@mit.edu}}

\abstract{
We present a brief report on a thought experiment in which an incident energetic parton traverses a brick of quark-gluon plasma (QGP), see Ref.~\cite{DEramo:2018eoy} for the full report.
We calculate the probability of detecting a parton showing up at a large angle with respect to its initial direction due to scattering with the constituents of QGP, using leading order perturbative QCD.  
We include all relevant channels, including the Rutherford-like channel as considered in early works, and those that are not Rutherford-like but become important at a large angle.
The resulting probability distributions contain information about the short distance structure of QGP. 
Our results provide key theoretical input toward finding the scatterers within the QGP liquid, which in turn is the necessary first step toward using precise, high-statistics, suitably differential measurements of jet modification in heavy ion collisions to study the evolution of the properties of QGP with changing resolution scale.
          }

\FullConference{International Conference on Hard and Electromagnetic Probes of High-Energy Nuclear Collisions\\
		30 September - 5 October 2018\\
		Aix-Les-Bains, Savoie, France}

\begin{document}

\section{Introduction}

%
%
%
\begin{figure} 
\center
\vspace{-0.3in}
\includegraphics[width=0.45\textwidth]{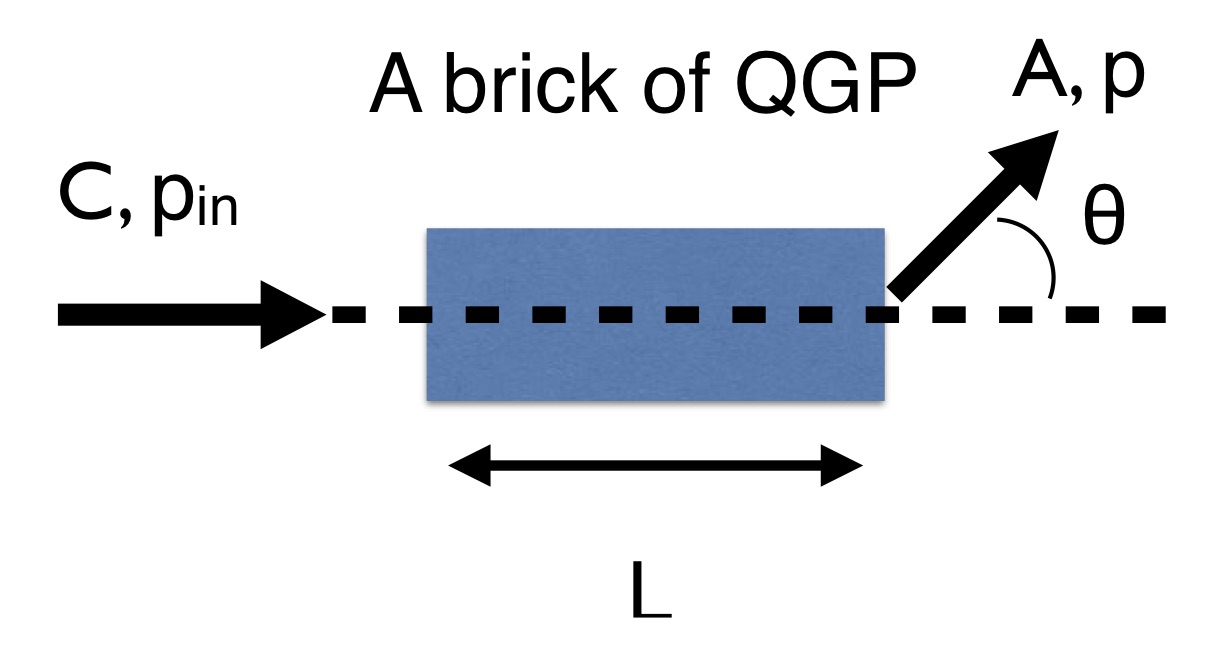}
\vspace{-0.1in}
\caption{
\label{fig:kinematic}
The set-up of the thought experiment that we analyze. 
We shoot an incident parton of ``type'' $C$ (type meaning gluon or quark or antiquark) with energy $p_{\rm in}$ through a ``brick'' of QGP with temperature $T$ and thickness $L$.   
An outgoing parton of type $A$ with energy $p$ is detected at an angle $\theta$ relative to the direction of the incident parton.  We shall calculate the probability distribution of $p$ and $\theta$ for a given $p_{\rm in}$. 
  }
 \end{figure}
 %
 %
%

The past decade has seen significant progress on understanding the behavior of quark-gluon plasma (QGP) at length scales longer than its inverse temperature $1/T$ (and consequently longer than its Debye length). 
Seen at this resolution,  
quarks and gluons behave collectively in QGP and form a ``near perfect'' liquid. 
 When seen at high resolution, on short length scales
quarks and gluons in QGP should look like weakly interacting quasi-particles because QCD is asymptotically free.
Understanding the evolution of the properties of QGP as a function of resolution scale is one of the important goals of heavy-ion collision experiments for the coming years~\cite{Geesaman:2015fha}.

Energetic parton showers are produced in the same collision as the droplet of QGP itself, and resolve the short distance structure and inner workings of QGP.
New and more precise jet data are anticipated in the 2020s, at RHIC from a state-of-the-art jet detector, called sPHENIX, and at the LHC from higher luminosity running.
Extracting quantitative information about short-distance structures of QGP from high precision data requires sophisticated modeling. 
In Ref.~\cite{DEramo:2018eoy}, 
we provide a key theoretical input for future phenomenological analysis. 
This development has also provided useful guidance on what sorts of observables experimentalists should aim to measure.

In Ref.~\cite{DEramo:2018eoy}, 
we set up a thought experiment as illustrated in Fig.~\ref{fig:kinematic}. 
We ``shoot'' an energetic quark or gluon of energy $\pinit$ through a static ``brick'' of QGP of thickness $L$ at a constant temperature $T$.
We use a very simplified model of QGP, viewing this ``brick'' just as a cloud of noninteracting massless quarks and gluons, which we refer to as medium partons. 
Due to the scattering between the incident parton and the medium partons,  
outgoing partons would show up with energy $p$ which can be different from $\pinit$, and with a nonzero angle $\theta$ relative to the direction of the incident parton. 
The object of our study is the phase space probability distribution $F(p, \theta)$, which contains rich information about QGP. 
We shall focus on the behavior of $F(p,\theta)$ at large $\theta$. 
In this region, a single scattering with large energy-momentum transfer dominates. 
Therefore the large $\theta$ behavior of $F(p,\theta)$  carries the information about the short distance structure of the QGP.  What we provide is only an input to future phenomenology, not quantitative predictions; in future, deviations between predictions and data at varying angles will contain information about how QGP resolved at varying scales deviates from a cloud of partons.

We will perform a standard leading order perturbative QCD calculation for $F(p,\theta)$. 
This is appropriate whenever the energy/momentum transfer is large. 
We have included all relevant binary scattering channels. 
For small $\theta$, 
``Rutherford-like scattering'' dominates~\cite{DEramo:2012uzl,Kurkela:2014tla}.
%
In our work, we consider $F(p,\theta)$ with generic $\theta$ and $p$. 
As we shall see shortly, a new feature emerges.

%

\section{Probability distribution $F(p,\theta)$}

\begin{figure} 
\centering
\includegraphics[height=0.24\textwidth]{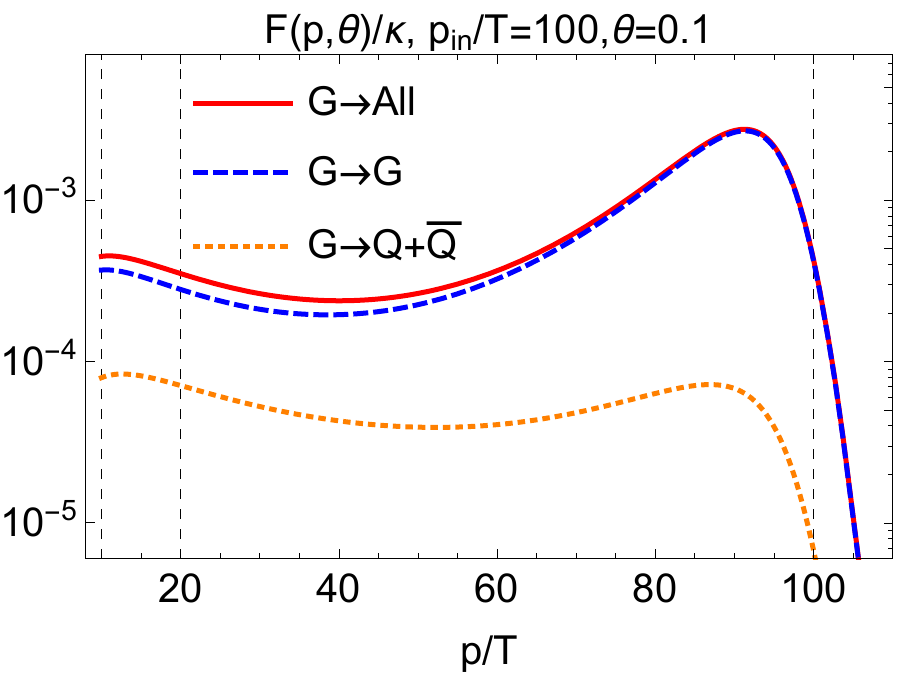}
\includegraphics[height=0.24\textwidth]{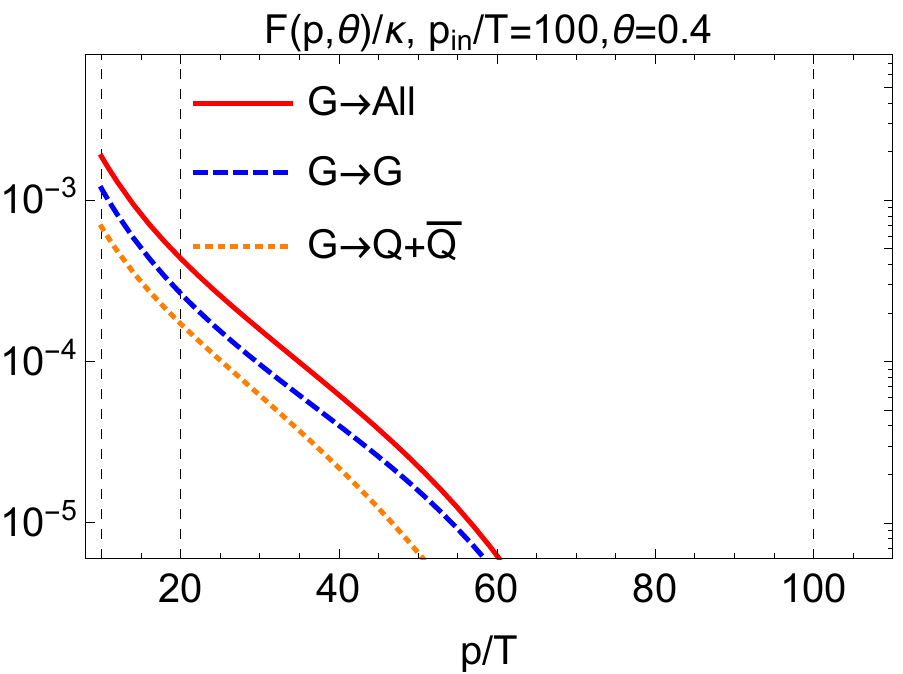}
\includegraphics[height=0.24\textwidth]{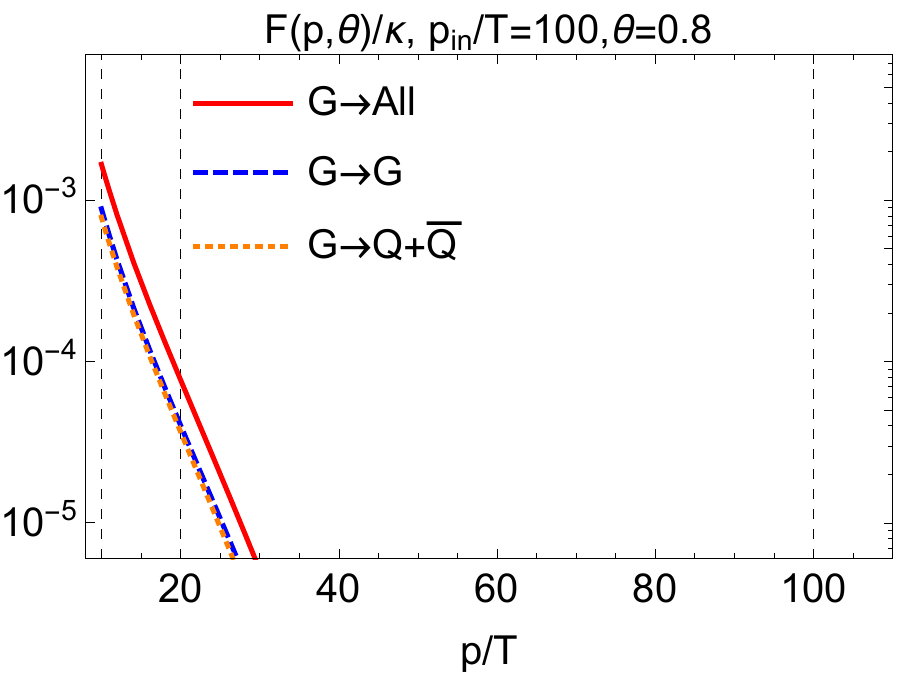}
\caption{
\label{fig:F}
The probability distributions $F\(p,\theta\)$ divided by $\kappa\equiv g^{4} T \CT$ plotted as functions of $\theta$ for an incident gluon with $\pinit/T=100$. From left to right, the columns correspond to choosing $\theta=0.1$, 0.2, 0.4.
The red solid, blue dashed and orange dotted curves show results corresponding to detecting an outgoing parton, or gluon, or quark (anti-quark) respectively. 
  }
\end{figure}

In this short proceeding,
we will only show and briefly discuss selected results with the incident parton being a gluon, 
see Ref.~\cite{DEramo:2018eoy} for a more comprehensive study. 
In Fig.~\ref{fig:F}, we present $F(p,\theta)$ as a function of $p$ for three different values of $\theta$, resulting from shooting a gluon with $\pinit=100 T$. 
In these plots, $F$ is divided by the dimensionless constant $\kappa\equiv g^4_{s} \CT T$ so that the rescaled results neither depend on the strong coupling constant $g_{s}$ nor on the duration $\CT=L/c$ this incident gluon spends in the QGP brick. 
For $\theta=0.1$, 
we observe that $F(p,\theta)$ features a peak at $p\approx \pinit$. 
This is expected as an outgoing parton with a very small angle is likely to have a small energy transfer $\pinit-p$. 
In contrast, at both $\theta=0.4$ and $\theta=0.8$, 
$F(p,\theta)$ decreases monotonically with increasing $p$. 
There are multiple layers of physics in this qualitative change in $p$-dependence as we increase $\theta$.
First, this is a manifestation of energy-momentum conservation. 
A larger $\theta$ means a larger momentum transfer in a single scattering, 
and such a scattering is more likely to be accompanied by a large energy transfer, which corresponds to a smaller $p$. 
Second, 
in this kinematic regime, outgoing partons often come from the scatterers in the QGP brick; one detects the ``kickee'' rather than the ``kicker''. 

To  demonstrate the latter point further, 
we plot the separate contributions due to detecting an outgoing gluon or an outgoing quark (or anti-quark) to $F(p,\theta)$ in Fig.~\ref{fig:F}. 
While at the smaller $\theta$, the outgoing partons are most likely to be gluons, the same species as the incident parton, finding an outgoing quark becomes equally probable at large values of $\theta$.
Since the incident parton is a gluon, 
this observation convincingly demonstrates that those outgoing quarks are coming from the medium quarks struck by the incident gluon.
This indicates that it is important from both qualitative and quantitative point of view to include all $2\to 2$ scattering processes in large angle regime. 
In addition, 
looking at $F(p,\theta)$ at larger values of $\theta$ can potentially tell us more information about the distribution of scatterers in the medium.

%


\section{Estimates based on phenomenologically motivated inputs}

\begin{figure} 
\centering
\includegraphics[height=0.24\textwidth]{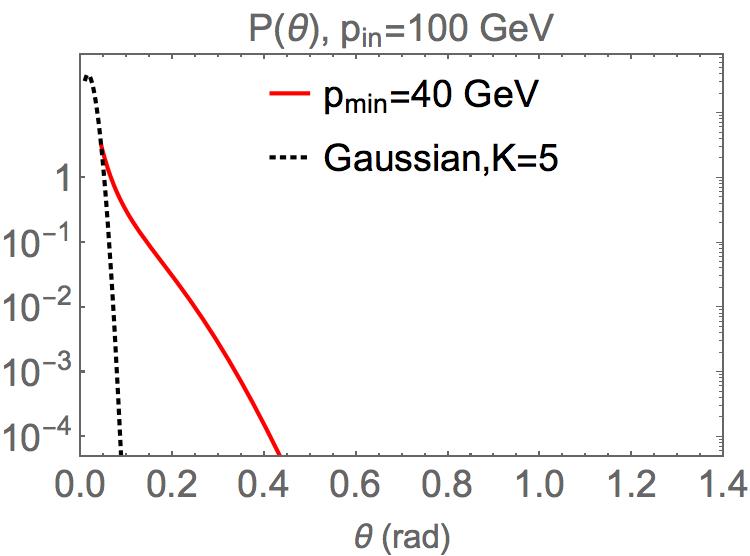}
\includegraphics[height=0.24\textwidth]{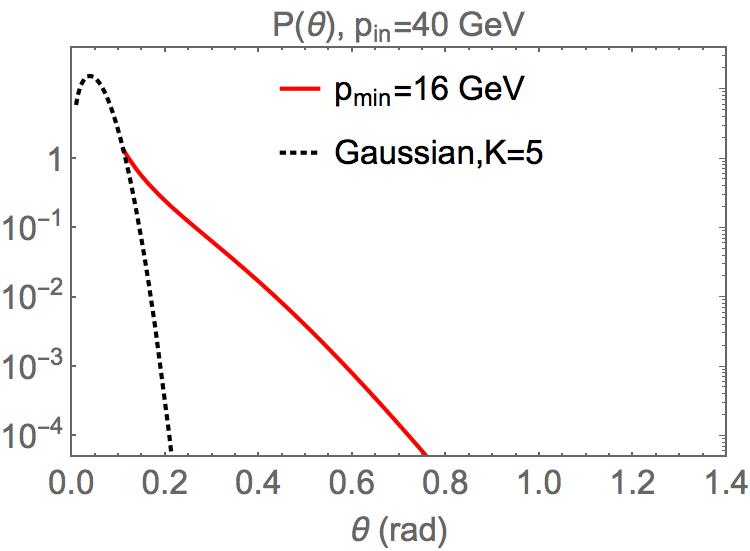}
\includegraphics[height=0.24\textwidth]{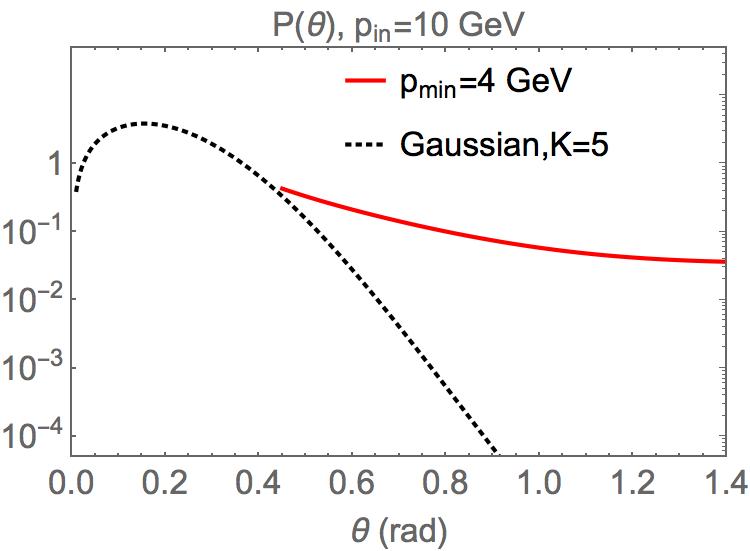}
\caption{
\label{fig:Pcompare}
We plot the angle distribution $P(\theta)$ (red solid curve) for an incident gluon.
For comparison, we plot the Gaussian angle distribution due to multiple soft scattering $P^{\rm{GA}}(\theta)$ (see text) as the dotted black curve.  
The ratio $p_{{\rm min}}/\pinit=0.4$ is fixed. 
From left to right, the columns correspond to using $\pinit=100, 40, 10$~GeV respectively.
  }
\end{figure}

In this section, we show the angle distribution $P(\theta)$ obtained by integrating $F(p,\theta)$ over $p$: 
$P(\theta)\equiv \int^{\infty}_{p_{\min}}\, dp\, F(p,\theta)\, .$ 
We will evaluate $P(\theta)$ by picking $T=0.4$~GeV as the temperature, and $\CT=3$~fm as the time that a parton spends in our brick of QGP. 
Since we are interested in binary collisions with characteristic energy transfer which is of the order $10$~GeV, 
we choose $g_{s}=1.5$ as the benchmark value for the strong coupling constant. 
By showing $P(\theta)$ with those phenomenologically motivated inputs, we can get a sense of how large $\theta$ should be so that a single scattering is dominant over multiple scatterings. 
In addition, the resulting $P(\theta)$ also give us a sense of how rarely single scattering process occur. 
For the sake of comparison, 
we will assume that when multiple scatterings become dominant, 
the angle distribution $P^{\rm{GA}}(\theta)$ would correspond to a Gaussian distribution in transverse momentum $p\sin\theta$. 
We parametrize the width of this Gaussian distribution through its connection to the jet quenching parameter $\hat{q}$, and pick a value of $\hat{q}$ guided by Ref.~\cite{Burke:2013yra} , see Ref.~\cite{DEramo:2018eoy} for more details. 

In Fig.~\ref{fig:Pcompare},
we present both $P(\theta)$ and $P^{\rm{GA}}(\theta)$ for three different energies $\pinit=100$, 40, 10~GeV. 
$P(\theta)$ depends on the lower bound $p_{\min}$ in  the computation 
of $P(\theta)$. 
To be specific, we fix the ratio $p_{\min}/\pinit=0.4$ to obtain results shown in Fig.~\ref{fig:Pcompare}. 
Plotting $P(\theta)$ and $P^{\rm{GA}}(\theta)$ together also gives us an idea on how the angle distribution may behave in a more complete calculation, where it should be dominated by single scattering as in our calculation at large enough angles but should be Gaussian at small momentum transfer. 
We observe that single scattering is dominant over multiple scatterings at a relatively small angle for a relatively large $\pinit$, e.g. $\pinit=100$~GeV. 
However, the magnitude of $P(\theta)$ drops very quickly with increasing $\theta$, making it difficult to measure experimentally. 
The magnitude of $P(\theta)$ can be significant in the single scattering dominant regime for $\pinit=10$~GeV, but the corresponding energy of outgoing partons is also soft. 
In this case, separating the outgoing partons from the medium partons will be very challenging.
The prospect is much more promising if we consider, for example, $\pinit=40$~GeV. 
Here,  the single scattering starts dominating over multiple scatterings at $\theta\approx 0.2$ and at the same time, 
the magnitude of $P(\theta)$ is still sizable. 

%
%
%
\begin{figure}[t] 
\center
\includegraphics[width=0.40\textwidth]{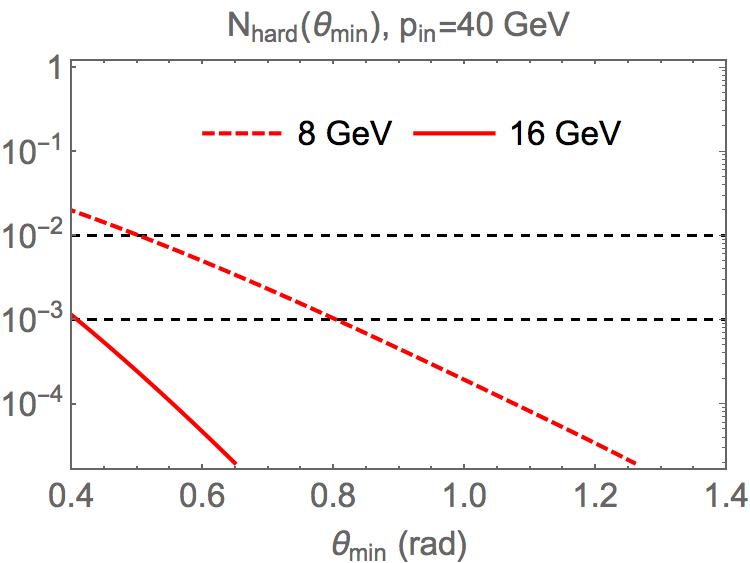}
\caption{
\label{fig:N}
$N_{\hard}(\theta_{\min})$ (see text) for an incident gluon with $\pinit/T=100$ 
The solid red curves (dashed red curves) show our results when we include
all partons with $p>p_{\min}=16$~GeV ($8$~GeV). 
The horizontal lines correspond to $N_{\hard}(\theta_{\min})=0.01$ and $0.001$.
  }
 \end{figure}
%
%
%

We finally show $N(\theta_{\rm{min}})$ in Fig.~\ref{fig:N}, the number of hard partons scattered by
an angle $\theta>\theta_{\min}$, namely
$N_{\rm hard}\(\theta_{\min}\)\equiv \int^{\pi}_{\theta_{\min}} d\theta\, \int^{\infty}_{p_{\min}} dp\, F\(p,\theta \)\, .$
Following the proceeding discussion, 
we consider an incident gluon with $\pinit=40$~GeV.
Reading Fig.~\ref{fig:N}, 
we see that the probability of seeing partons with energies greater than $8$~GeV at angles $\theta > 0.5$ is $1/100$ and
at angles $\theta>0.8$ is $1/1000$. 
Although processes related to large angle scattering are rare, the relevant probabilities are not tiny, given the anticipated high statistics data sets for jets in the not too distant future.

\section{Summary and outlook}

We have conducted the thought experiment depicted in Fig.~\ref{fig:kinematic} and computed the probability distribution $F(p,\theta)$ for finding a parton subsequently with an energy $p$ that has been scattered by an angle $\theta$ relative to the direction of the incident parton.
We furthermore obtained  the angle distribution $P(\theta)$ as well as $N_{\rm hard}(\theta_{\min})$, the number of outgoing parton with angle $\theta>\theta_{\min}$.
We considered binary collisions in which the incident parton strikes a single parton from the medium. 

In future quantitative studies, we should replace the single incident parton considered here with a shower of energetic partons described via a jet Monte Carlo code, and replace a static brick of QGP with an expanding cooling droplet of QGP as created in heavy-ion collisions.  With a realistic phenomenological simulation in place, the challenge will be to analyze jet substructure observables
that can answer questions like whether jets containing a $\sim 40$~GeV parton sprout an extra prong corresponding to a $\gtrsim 8$~GeV parton scattered at $\theta > 0.8$ with probability $1/1000$.
Ultimately, 
high staistics jet measurements together with realistic phenomenological simulations can pave the way towards learning about the microscopic structure of liquid QGP.

%
%


\textbf{Acknowledgement}-- This work was supported in part by the Office of Nuclear Physics of the U.S. DOE under Contract Number~DE-SC0011090~(KR, YY) and by Istituto Nazionale di Fisica Nucleare (INFN) through the ``Theoretical Astroparticle Physics'' (TAsP) project~(FD).

\end{document}